\documentclass[a4paper]{spie}  %>>> use this instead for A4 paper
\usepackage{amsmath,amsfonts,amssymb}
\usepackage{graphicx}
\usepackage{caption}
\usepackage[colorlinks=true, allcolors=blue]{hyperref}

\title{Towards a physical understanding of the thermal background in large ground-based telescopes}

\author[a]{Leonard~Burtscher}
\author[a]{Ioannis~Politopoulos}
\author[b]{Sergio~Fern\'andez-Acosta}

\author[c]{Tibor Agocs}
\author[d]{Mario~van~den~Ancker}
\author[e]{Roy~van~Boekel}
\author[a,f]{Bernhard~Brandl}
\author[d]{Hans Ulrich K\"aufl}
\author[g]{Eric~Pantin}
\author[h]{Alex~G.~M.~Pietrow}
\author[d]{Ralf Siebenmorgen}
\author[a]{Remko Stuik}
\author[i]{Konrad~R.~W.~Tristram}
\author[i]{Willem-Jan~de~Wit}

\affil[a]{Leiden Observatory, Leiden University, P.O. Box 9513, 2300 RA Leiden, The Netherlands}
\affil[b]{GRANTECAN (GTC), Centro de Astrof\'isica de La Palma, Cuesta de San Jos\'e, 38712 Bre{\~n}a Baja, La Palma, Spain}
\affil[c]{NOVA O/IR Group, P.O. Box 2, 7990 AA Dwingeloo, The Netherlands}
\affil[d]{ESO, Karl-Schwarzschildstr.2, 85748 Garching, Germany}
\affil[e]{Max-Planck-Institut f\"ur Astronomie, K\"onigstuhl 17, 69217 Heidelberg, Germany}
\affil[f]{Faculty of Aerospace Engineering, Delft University of Technology, Kluyverweg 1, 2629 HS Delft, The Netherlands}
\affil[g]{AIM, CEA, CNRS, Universit\'e Paris-Saclay, Universit\'e Paris Diderot, Sorbonne Paris Cit\'e, F-91191 Gif-sur-Yvette, France}
\affil[h]{Institute for Solar Physics, Dept. of Astronomy, Stockholm University, Albanova University Centre, SE-106 91 Stockholm, Sweden}
\affil[i]{European Southern Observatory, Casilla 19001, Alonso de Cordova 3107 Vitacura, Santiago, Chile}

\authorinfo{Corresponding author e-mail address: burtscher@strw.leidenuniv.nl}

\begin{document} 
\maketitle

\begin{abstract} %250 words max.
Ground-based thermal-infrared observations have a unique scientific potential, but are also extremely challenging due to the need to accurately subtract the high thermal background. Since the established techniques of chopping and nodding need to be modified for observations with the future mid-infrared ELT imager and spectrograph (METIS), we investigate the sources of thermal background subtraction residuals.

Our aim is to either remove or at least minimise the need for nodding in order to increase the observing efficiency for METIS. To this end we need to improve our knowledge about the origin of chop residuals and devise observing methods to remove them most efficiently, i.e. with the slowest possible nodding frequency.

Thanks to dedicated observations with VLT/VISIR and GranTeCan/CanariCam, we have successfully traced the origin of three kinds of chopping residuals to (1) the entrance window, (2) the spiders and (3) other warm emitters in the pupil, in particular the VLT M3 mirror cell in its parking position. We conclude that, in order to keep chopping residuals stable over a long time (and therefore allow for slower nodding cycles), the pupil illumination needs to be kept constant, i.e. (imaging) observations should be performed in pupil-stabilised, rather than field-stabilised mode, with image de-rotation in the post-processing pipeline. This is now foreseen as the default observing concept for all METIS imaging modes.
\end{abstract}

% Include a list of (max. 8) keywords after the abstract 

%%
%% +-+-+-+-+-+-+-+-+-+-+-+-+-+-+-+-+-+-+-+-+-+-+-+-+-+-+-+-+-+-+-+-+-+-+-+-+-+
%%

\section{INTRODUCTION}
\label{sec:intro}

The thermal infrared\footnote{We use the term thermal infrared to refer to the wavelength range of $\sim$ 3 - 25 $\mu$m, where the thermal emission from the telescope is the dominant source of photons. This encompasses the atmospheric bands $L$, $M$, $N$, and $Q$. The former two are usually referred to as being part of the ``near-infrared'' part of the spectrum and the latter two are referred to as the ``mid-infrared''.} is the wavelength range of choice for observing highly obscured and warm objects such as active galactic nuclei or proto-planetary disks, and is increasingly attracting attention also for the characterisation of Earth-like exoplanets\footnote{For a recent summary of thermal-infrared science and technology, please see the \href{https://www.eso.org/sci/meetings/2020/IR2020.html}{proceedings (presentation PDFs and video recordings) of the on-line ESO workshop IR 2020}, held in October 2020\cite{burtscher2021b}.}. On the other hand, ground-based thermal infrared observations are plagued by a very high thermal background that needs to be removed before revealing the sources of astrophysical interest. 

For example, the point source sensitivity of the Mid-infrared ELT Imager and Spectrograph (METIS\cite{brandl2018}) in the red part of the atmospheric $N$ band (filter N2) is 147 $\mu$Jy under median conditions, but the background-equivalent flux in this filter is about 19 Jansky (Jy). This immediately shows how accurately the background needs to be removed -- to about 1 part in 400,000 -- in order to reach the Background LImited Performance (BLIP). Since the accuracy of the flat field correction for current detectors is not known to a $10^{-5}$ relative precision level, the observation of a faint source sunk into a large background has to be based on operational methods.

This removal is usually achieved through a combination of offsets following the idea that a detector pixel, which sees the source and the background, sees the same (or a similar) background if the offset is commanded on a timescale shorter than any variation in the background or detector response. 

\begin{figure} [ht]
\begin{center}
\includegraphics[width=10cm]{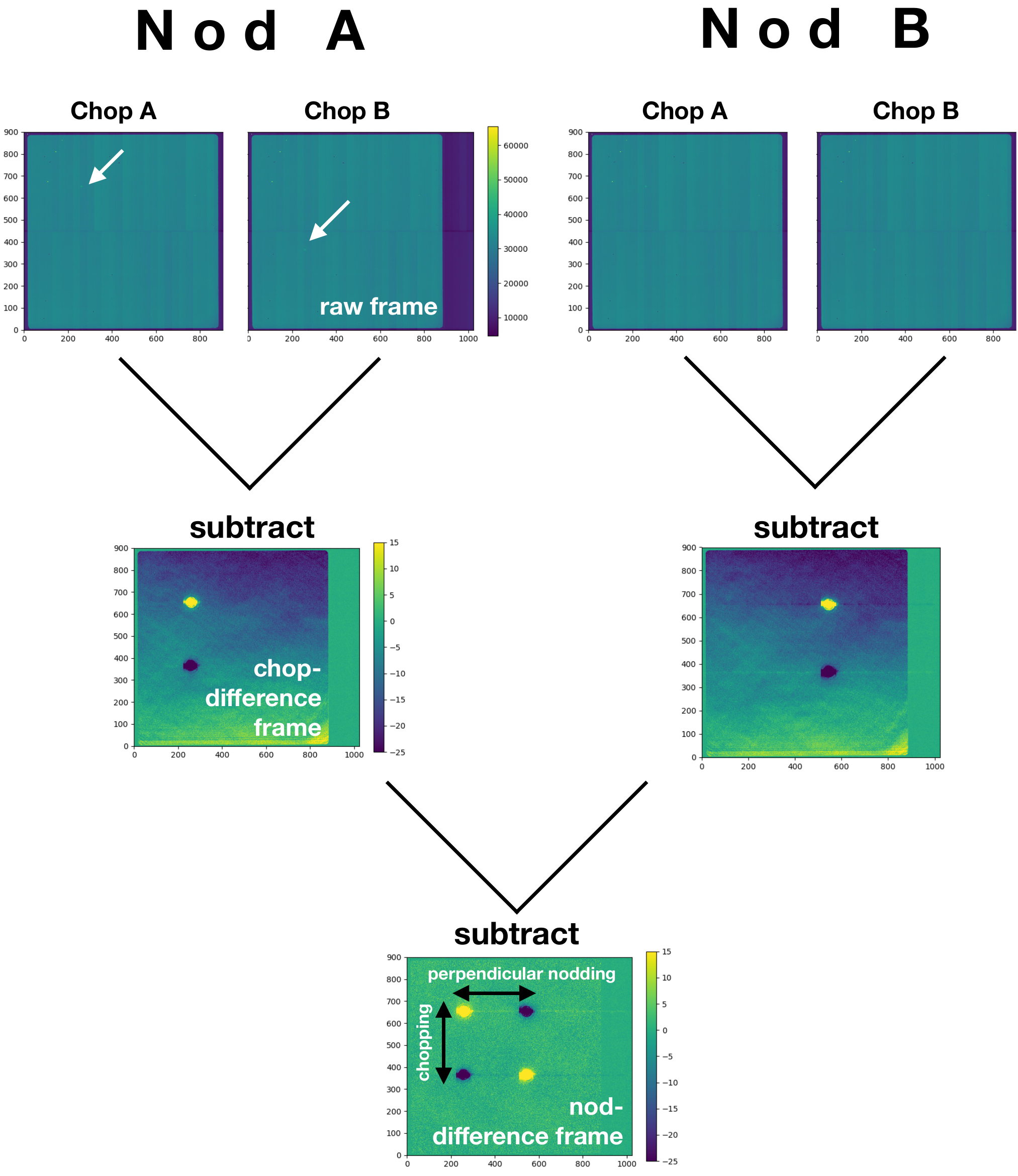}
\end{center}
\caption{\label{fig:chop_nod} The standard scheme of background subtraction in ground-based thermal-infrared imaging used at all major telescopes: Fast ($\sim$ 1\ldots10 Hz) offsets with the secondary mirror (``chopping'') are used to subtract detector and atmospheric fluctuations. The residuals from chopping are removed by offsetting the telescope as a whole (``nodding'', here specifically ``perpendicular nodding'') and repeating the chopping there. The combination of these four positions results in a near-perfect background subtraction, but at significant operational cost.}
\end{figure}

The default background subtraction strategy in ground-based thermal-infrared astronomy has therefore been, since a long time, chopping and nodding. Chopping alternates the beam between an on-axis and a nearby off-axis position (typical offsets are 5--30 arc seconds) on a timescale of a second or faster in order to subtract {\em fast} variations in the observed thermal background. Since the off-axis position sees a slightly different optical path through the telescope, {\em chop residuals} (or {\em thermal offsets}) remain. These slowly varying residuals can be removed by {\em nodding}, i.e. by alternating the {\em telescope} position on an $\approx$ minute timescale and observing the same chopping pattern there. In this way, ideally, the same {\em chop residuals} are observed again and can be removed. The nodding direction is usually chosen to be perpendicular to the chop direction ({\em perpendicular nodding}) or identical to it ({\em parallel nodding}). In parallel nodding only three beams remain after the chop/nod subtraction, resulting in an $\approx$ 15\% higher Signal to Noise Ratio (SNR). Parallel nodding may come at a cost of image quality however when nodding and chopping offsets are not exactly identical. As a result the central beam can be smeared out. The standard mode of VISIR \cite{VISIR-manual} is therefore perpendicular nodding and is illustrated in Fig.~\ref{fig:chop_nod} using real VLT/VISIR data. In METIS, we expect the chopping and nodding offsets to be very accurate thanks to Adaptive Optics (AO) guiding and our internal chopper, so we can make use of the improved SNR of parallel nodding.

The origin and fluctuation timescales of the thermal ``background'' (which is actually a ``foreground'') are not very well understood despite a long history of research on the topic \cite{kaeufl1991, mason2008, pantin2010}. This is partly to blame on the fact that thermal-IR detectors are typically not stable enough on timescales of seconds making it difficult to disentangle detector residuals from other residuals originating outside the instrument. We know, however, at least three components that contribute to these variations: the sky, the telescope and the detector. From earlier experiments with VISIR\cite{pietrow2016} (see also below), we know that the chopping residuals can in some cases be {\em very} stable over timescales as long as {\em months}. Unfortunately, there is only little information available over the actual fluctuation timescales of the thermal-infrared sky. An older study by H.-U. K\"aufl from 1991 \cite{kaeufl1991} suggests that chopping at $\approx$ 8 Hz is required to remove sky background fluctuations, but more recent results \cite{miyata1999,pantin2010} suggest that the sky background itself does not vary on such fast timescales (see Fig.~1.5 in E. Pantin's habilitation thesis\cite{pantin2010}) and chopping at $\gtrsim 1$ Hz should be sufficient to remove the {\em sky} fluctuations. This is in agreement with a ``drift scanning'' experiment with the ``old'' VISIR instrument (when it was equipped with a DRS detector) where a chopping frequency of $\approx$ 1 Hz was sufficient to reach the background-limited performance in all filters\cite{ives2014}.

\subsection{ELT/METIS}
The Mid-Infrared ELT Imager and Spectrograph (METIS), a first-generation instrument for the ELT expected on-sky in 2028, will be a major advancement in ground-based thermal-infrared astronomy. It will make full use of the ELT's unprecedented sharpness and light collecting power that is particularly powerful in background-limited conditions when the integration time $t_{\rm int}$ to reach a given point-source sensitivity scales with telescope diameter $D$ as $t_{\rm int} \propto D^{-4}$. Thanks to METIS' versatile observing modes, it will offer diffraction-limited imaging and long-slit spectroscopy in the atmospheric $L$, $M$ and $N$ bands as well as high-resolution integral-field unit spectroscopy in the $L$ and $M$ bands; most modes can be combined with coronagraphy for high-contrast imaging.

One of the biggest challenges that we expect for thermal-infrared observations at the ELT is to accurately subtract the thermal background of the telescope and the sky such that METIS can reach its full sensitivity as a thermal-background limited instrument. For this we do not only need to subtract the thermal background much more accurately than currently (to reach $\mu$Jy sensitivities), but we also need to employ novel background subtraction techniques since the standard technique of chopping with M2 (see Fig.~\ref{fig:chop_nod}) is not possible at the ELT. Already the VLT's $\approx$ 1.1m diameter M2 unit was specifically designed to be light (e.g. the mirror is made of Beryllium), reaching a mass of just 1800 kg allowing for chopping frequencies of up to 5 Hz. % https://www.eso.org/sci/facilities/paranal/telescopes/ut/m2unit.html
The 4.2m-diameter M2 at the ELT, however, weighs 12 tonnes, and cannot be used for chopping. %https://www.eso.org/sci/facilities/eelt/telescope/mirrors/
While the tip-tilt mirror (M5 in the ELT's optical train) could in principle be used for chopping, its stroke is required for field stabilisation.
The METIS project has therefore decided early on to commission the design of a cryogenic built-in chopper\cite{paalvast2014}. The METIS chopper is designed to chop at frequencies up to 10 Hz which was necessary to remove the so called Excess Low-Frequency Noise (ELFN) of the originally foreseen detector AQUARIUS\cite{ives2014}. The recent development of a new HgCdTe-based detector (``GeoSnap'') for the thermal-infrared domain does not seem to suffer from this noise, but still requires chopping at frequencies of $\approx$ 1 Hz to remove 1/f noise. The origin of this particular noise source in the GeoSnap detector is so far unknown.\cite{atkinson2020}

\begin{figure} [ht]
\begin{center}
\includegraphics[width=10cm]{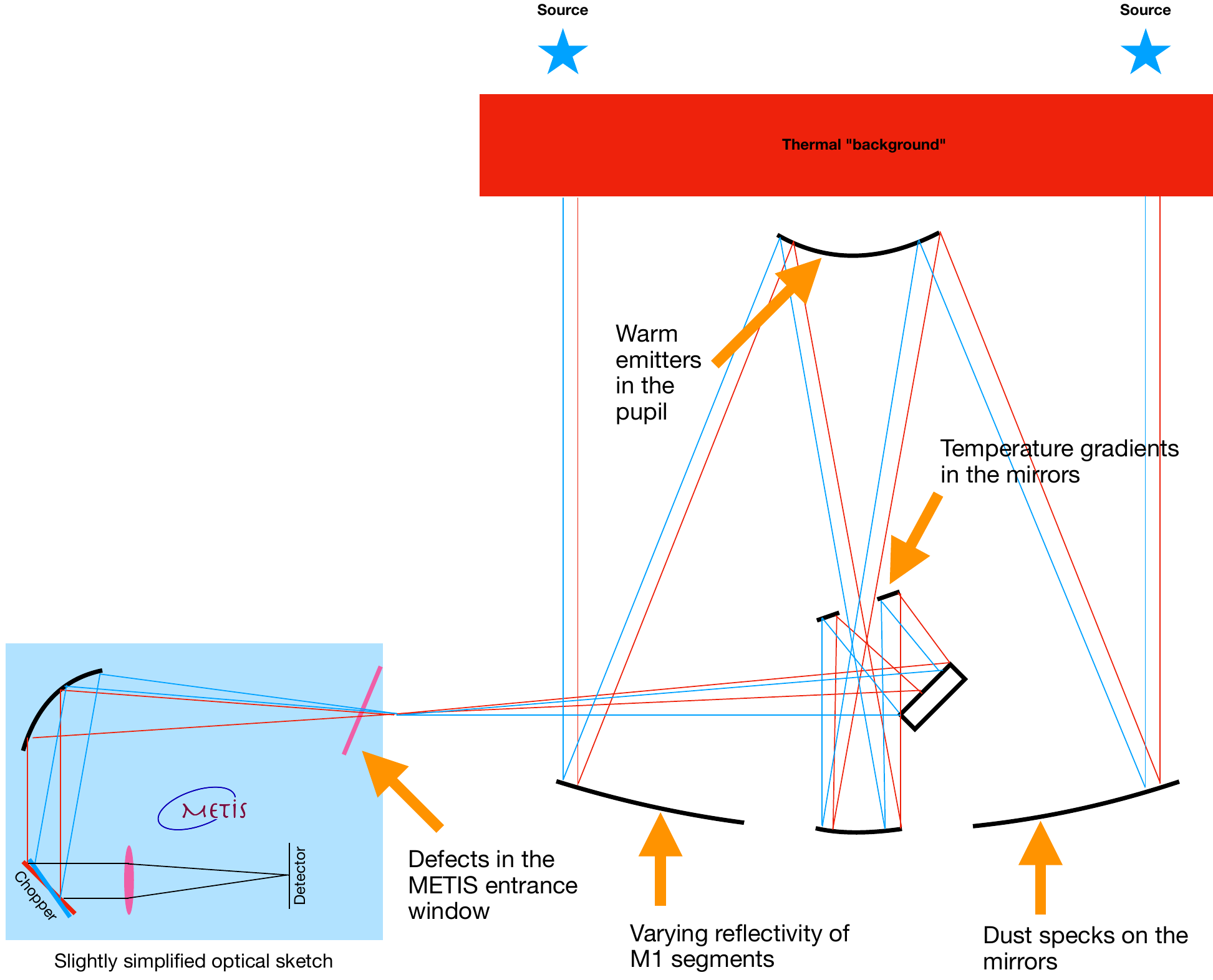}
\end{center}
\caption{\label{fig:chopping_ELT} Challenges of chopping with a cryogenic chopper like in ELT/METIS: An offset field (e.g. an empty patch of sky) does not follow the same path (red lines) through the warm optics as the source (blue lines). When the chopper is used to rapidly switch between the source and sky regions, the thermal background (which is dominated by the warm optics) changes and this in turn leads to imperfect background subtraction with chopping. A variety of possible effects for these slight variations in the background signal are indicated in the cartoon. Nodding removes these chop residuals, but at the expense of operational overheads. Objects are not to scale and angles/offsets are exaggerated for clarity.}
\end{figure}

The METIS chopper is located in the pupil plane of the second re-imager within METIS. It is the 11th mirror inside METIS and the upstream optical path includes the (also cryogenic) single-conjugate adaptive optics pickoff dichroic as well as the cryostat's entrance window, and the six warm mirrors of the ELT (five from the telescope and one on the ELT Nasmyth pre-focal station). In further contrast to existing telescopes, the ELT contains two further upwards facing mirrors besides M1, i.e. M3 and M5, that will also require regular cleaning. Since chopping in METIS is performed so far downstream in the optical path, emissivity inhomogeneities or surface defects will play a role in a number of locations (see Fig.~\ref{fig:chopping_ELT} for an illustrative cartoon). A simple ``chop-difference frame'' will therefore probably look more complex than in present-day ground-based thermal-IR instruments like VISIR (see Fig.~\ref{fig:chop_nod}).

In addition, the emissivity of the ELT will be larger than the VLT's due to the larger number of mirrors. The total emissivity of the VLT/VISIR system (including M1, M2, M3, the spiders and the VISIR entrance window) can be estimated to be about 9 \% (own calculations based on observations\footnote{\href{https://github.com/astroleo/visir_sensitivity/blob/master/VISIR\%20sensitivity.ipynb}{See this Python notebook for a derivation.}}). For METIS at the ELT we expect the effective emissivity of the telescope and the METIS entrance window to be $\sim$ 19\%. This number includes the clean mirror emissivity (4\%), segment gaps (1\%), and an estimated average level of dust on the telescopes (12\%)\footnote{We note that this is a conservative estimate assuming that the emissivity can be computed from 1 - reflectivity. This is not actually true since the loss of reflectivity due to dust leads to a large part to scattering rather than added emissivity\cite{holzloehner2018}. Most of the scattering is near the specular reflection and therefore does not lead to significantly added emissivity since radiation from the cold sky is scattered into the field.}\cite{holzloehner2018}. In addition, the emissivity from the ZnSe substrate of the METIS entrance window contributes $\approx$ 2\% to the total emissivity at 10 $\mu$m (as per the METIS transmission budget).

The sky at the ELT, on the other hand, is only marginally colder due to the slightly higher elevation of the ELT compared to the VLT. Compared to the VLT/VISIR system, the telescope will be more dominant than the sky background at almost all wavelengths for ELT/METIS.

\subsection{Novel background removal techniques}

Chopping and nodding are in principle very simple techniques, but nodding comes with significant operational demands and overheads, especially with METIS at the ELT which will be AO-guided in all observing modes. If the same nodding frequency was chosen for METIS as for VISIR (typically 90s for one cycle\cite{VISIR-manual}), and if the re-acquisition (closing the AO loop on the guide star) takes 10 seconds\cite{ELT-common-ICD}, then nodding would reduce the observing efficiency by more than 20\%\footnote{There is yet another possible issue with slow nodding at the ELT, however, since the so called Low-Order Optimisation regularly changes the shape of the primary mirror. We note that VISIR never experienced changes in the thermal background structure during Active Optics optimisation (priv. comm. ESO VISIR Instrument and Operations Team), but in case we choose a chop/nod scheme with long nodding cycles, this is an issue that will require careful observation.}. Chopping with the internal cryo-chopper in METIS only leads to a minor inefficiency, on the other hand. At 1 Hz chopping, i.e. 2 offsets per second, and a detector integration time of 10 ms, only two frames per second would have to be removed to wait for the chopper to settle. This corresponds to an observing efficiency of 98 \%. Chopping without nodding is not an option, however, as it would lead to significantly degraded performance\cite{pantin2015}. It is therefore clear that understanding the chopping residuals holds great promises for increased observing efficiency, if it can reduce the nodding frequency.

Already in 1992, Landau et al.\cite{landau1992} demonstrated that three beam chopping can improve the background subtraction with a single-pixel detector. Recently, using test data on VLT/VISIR, we showed that a similar technique can achieve essentially the same signal/noise ratio as traditional chopping/nodding on the VLT \cite{pietrow2016,pietrow2019}. Due to limitations in the VLT M2 mirror control, we could not test a three-point chopping method as Landau's\cite{landau1992}, but we could alternate on $\approx$ minute timescales between standard observations chopped North and observations chopped South. Combined, this ``Inverse Chop Addition'' resulted in a background subtraction that is on par with the standard chop/nod method, at least for short observations. We believe that this is currently one of the most promising non-standard background-removal techniques applicable for METIS, and our Vortex coronagraph for $N$ band high contrast imaging has been specifically designed to support it.

A second important conclusion from this work was that the chopping residuals in short observations where the telescope was not tracking seemed to stay constant over much longer times than the normal nodding cycle (90 seconds), suggesting that the dominant structure for the chop residuals is instrumental or from the telescope, rather than from sky fluctuations.

Other novel background removal techniques include drift scanning that can at least be used to improve the flat field calibration\cite{heikamp2014}, but possibly also for general background subtraction\cite{ohsawa2018}. The CanariCam team performed further experiments with drift scanning as a method for background subtraction in ELTs; early results from these tests will be presented in Packham, Torres-Quijano, and Fernandez-Acosta in these SPIE proceedings.

A completely different approach for sky-subtraction in the thermal-infrared $L$ and $M$ bands has been presented by Marois et al. 2010\cite{marois2010} and may also be applicable to the N band. Their approach, Locally Optimized Combination of Images (LOCI), is similar to Principal Component Analysis (PCA): the authors removed the sky background from staring frames without chopping. Hunziker et al.\cite{hunziker2018} further showed that this method can be used to achieve the BLIP in L band high-contrast imaging (HCI) observations and a similar PCA-based approach has been part of the HCI data reduction package VIP\cite{gonzalez2017} since 2015.

Chopping has always been problematic in crowded fields (e.g. the Galactic Center) where even the maximum chopping offsets that M2 chopping allows are not sufficient to find a field free of sources. To overcome these limitations, the introduction of a ``Dicke switch'' has been suggested and successfully deployed in the NEAR experiment \cite{kasper2020}. In such a setup, the optical path that the detector sees is alternated quickly between the sky and an internal reference source whose emissivity is matched to the sky in order to allow very precise detector calibration. This comes at the cost of reducing the observing efficiency by at least a factor of two, however. An alternative approach to observe extended, crowded fields has been presented by Robberto et al. (2005)\cite{robberto2005} who used image reconstruction techniques based on a minimisation of the negative chop residuals from real sources.

%%
%% +-+-+-+-+-+-+-+-+-+-+-+-+-+-+-+-+-+-+-+-+-+-+-+-+-+-+-+-+-+-+-+-+-+-+-+-+-+
%%

\section{OBSERVATIONS}
We continued our investigation into chop-difference frames of VISIR with the ESO calibration programme 4101.L-0802. Observations took place in July 2018 in visitor mode. 0.3 nights were scheduled between 12 and 13 July and 0.4 nights between 17 and 18 July. Due to an overlap with MATISSE commissioning, some extra time could also be used in twilight on 10 and 11 July to test the foreseen experiments and templates. Thanks to the excellent support from Paranal staff, templates could be adapted to support non-standard modes or fixed (e.g. the spectroscopy burst mode template) to make it work on time. Visitor mode was essential for this programme to supervise the experiments and control or know exactly what the telescope and instrument did.

All foreseen tests were performed successfully. Thanks to the accessibility of twilight time on 10 and 11 July, some tests could even be repeated. Table~\ref{tab:observations} summarises all {\em successful} observations of this calibration programme. Note that pupil images, while useful for understanding the instrument, are not ingested into the ESO archive. These can be obtained by e-mailing the corresponding author.

\begin{table}[h]
   \centering
   \caption{Summary of all successful tests performed in the VISIR calibration programme 4101.L-0802.}
   \begin{tabular}{lllp{80mm}} % Column formatting, @{} suppresses leading/trailing space
      \hline
      id    & DATEOBS of first file & DATEOBS of last file & Description\\
      \hline
      1		& 2018-07-10T21:40:34 & 2018-07-10T21:54:58 & good VISIR pupil images: known filter (ARIII), in focus and not saturated \\
      2		& 2018-07-10T22:09:30 & 2018-07-10T22:46:38 & rotate ADA.ABSROT while taking chop/nod images \\
      3		& 2018-07-11T22:46:12 & 2018-07-11T22:54:01 & pupil image with closed Cassegrain shutter\\
      4		& 2018-07-11T22:56:31 & 2018-07-11T23:07:23 & burst mode image with closed Cassegrain shutter (for ELFN analysis, but only imaging, no spectroscopy, i.e. no power spectrum)\\
      5		& 2018-07-11T23:51:41 & 2018-07-11T23:52:39 & pupil images with dome closed\\
      6		& 2018-07-12T22:41:12 & 2018-07-12T23:13:41 & CHOP POSANG change experiment in three filters: ARIII, SIV\_1, PAH2\_2\\
      7		& 2018-07-12T23:16:09 & 2018-07-12T23:54:48 & ADA ABSROT change experiment in three filters: PAH2\_2, ARIII, SIV\_1 (use the 360 deg frame of the previous test as the beginning of this test)\\
      8		& 2018-07-13T00:46:35 & 2018-07-13T01:01:37 & observation of ups Lib during max. field rotation in pupil tracking mode\\
      9		& 2018-07-13T01:08:18 & 2018-07-13T01:34:07 & like (8) but with HD 142332\\
      10	& 2018-07-13T01:52:56 & 2018-07-13T02:01:27 & rotate ADA while taking chopped frames (movie)\\
      11	& 2018-07-13T02:16:12 & 2018-07-13T02:16:12 & like (10) but in pupil imaging\\
      12a	& 2018-07-18T06:02:28 & 2018-07-18T06:12:25 & observation of M30 in pupil-tracking mode before culmination\\
      13	& 2018-07-18T06:44:30 & 2018-07-18T07:00:12 & burst mode spectrum of the sky at three airmasses\\
      12b	& 2018-07-18T07:06:40 & 2018-07-18T07:44:05 & observation of M30 in pupil-tracking mode after culmination\\
      14	& 2018-07-18T07:59:36 & 2018-07-18T08:07:53 & Burst mode images of the sky in SIV\_1 filter at airmass 1.3\\
      15	& 2018-07-18T08:58:11 &	2018-07-18T10:14:43 & variation of ADA.ABSROT and CHOP.POSANG (a chopping cross at each ADA.ABSROT position)\\
      16	& 2018-07-18T10:21:06 & 2018-07-18T10:21:06 & spec burst mode with Cassegrain shutter closed\\
      \hline
   \end{tabular}
   \label{tab:observations}
\end{table}

We analysed all observations ``as is'', i.e. without any data processing except for the usual chop subtraction. Only for the pupil images did we perform a slightly different processing and removed the bias in each channel by subtracting the median counts from the region not illuminated by the pupil. To measure the raw counts from the thermal background in normal imaging observations, we subtracted the median detector bias as observed under the cold field mask at the edge of the detector.

In parallel, we also performed a number of tests with CanariCam. CanariCam\cite{telesco2003} is the facility multi-mode thermal-IR (8-25 $\mu$m) camera on the 10-m Gran Telescopio CANARIAS (GTC) at La Palma, Spain.  It provides imaging, spectroscopic and unique polarimetric capabilities at, or near, the diffraction limit of the telescope.
Since 2012, it had been operating in queue mode at one of the Nasmyth focal stations, until it was temporarily decommissioned in 2016.  Following an upgrade project\cite{fernandezacosta2020}, it was reinstalled and recommissioned in a folded-Cassegrain focal station in late 2019. Due to the demanding instrumentation programme at GTC, and the need to host the next generation instruments, CanariCam is scheduled to be decommissioned from the telescope in January 2021.

\begin{figure}[ht]
\centering
\captionof{table}{Log of the CanariCam observations. These engineering data are not available through the GTC archive, but can be requested from the authors. For rows 2 and following, configuration changes compared to the previous row are marked in red.}
\label{tab:observations_cc}
\includegraphics[width=\columnwidth]{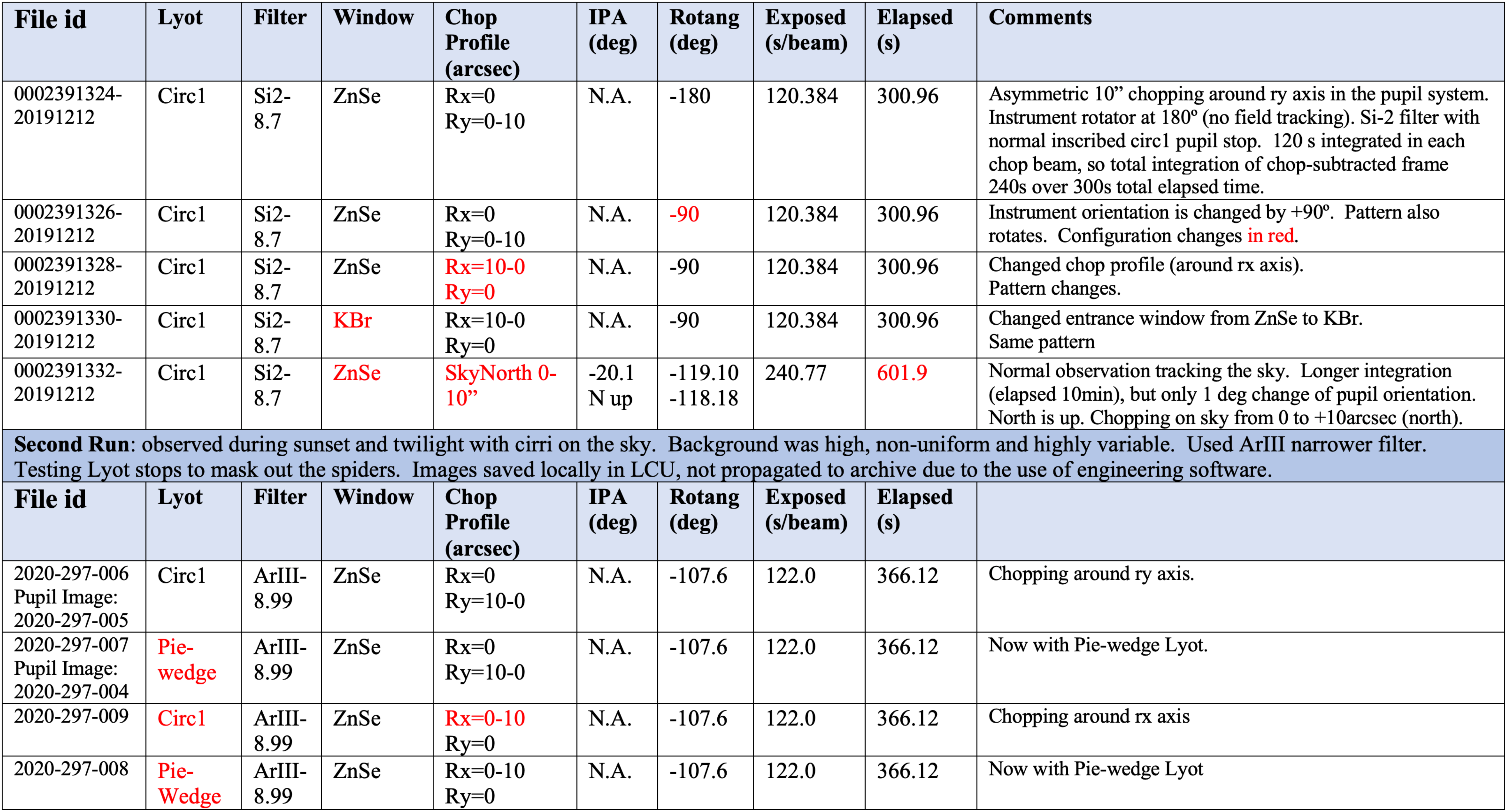}
\end{figure}

%%
%% +-+-+-+-+-+-+-+-+-+-+-+-+-+-+-+-+-+-+-+-+-+-+-+-+-+-+-+-+-+-+-+-+-+-+-+-+-+
%%

\section{RESULTS and ANALYSIS}
\subsection{Gradient and fringes in VISIR chop residual images}
As had already been noted in earlier data, one of the most noticeable features in VISIR chop-difference images is a significant gradient of variable strength and orientation. While this gradient would not necessarily present a problem for the scientific analysis by itself, we study it since it is a useful {\em tool} to understand the link between telescope properties and chopping residuals.

\begin{figure}[ht]
\centering
\includegraphics[width=\columnwidth]{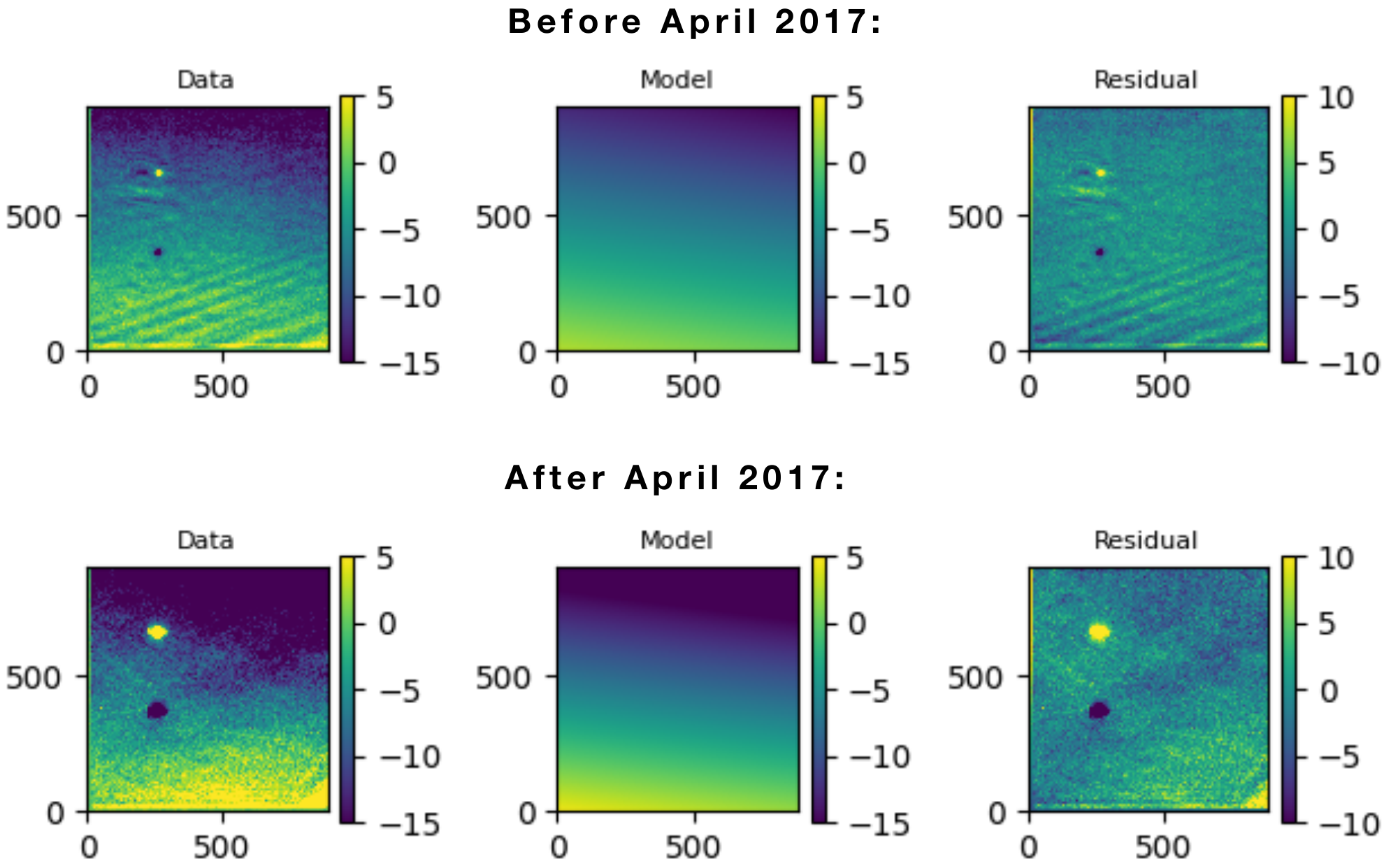}
\caption{\label{fig:chop_gradient_model_window}Chop-difference frames before (upper) and after (lower row) the exchange of the cryostat entrance window in April 2017. From left to right in each row: Chop-difference image, linear gradient model (middle), residual (right). The ripples/fringes in the lower right corner have disappeared since the window exchange.}
\end{figure}

In Fig.~\ref{fig:chop_gradient_model_window} we show typical examples of chop-difference images from before and after the exchange of the cryostat entrance window in April 2017. Apart from the gradient, it is immediately obvious that the structure of the chop residuals changed. In particular the fringes in the lower right corner of the detector and the ``crater'' close to the positive image of the star in this particular observation, are gone since the window exchange. Since they always appeared in the same position in the detector (and did not rotate like the telescope pupil), they must have been part of the instrument. Although obvious tests for fringing, such as a change of the pattern for different filters, have been negative, we conclude that they must have originated in the old entrance window.

The gradient itself can easily be fitted with a linear model, although in a few observations a quadratic model returns a significantly better residual. Since the gradient is only marginally detected in some images, we will use the simpler linear gradient model in the following, however.

\subsubsection{VISIR gradient behaviour under instrument rotation}
To locate the origin of the gradient, we are interested in its dependency on the rotation of the instrument using the Cassegrain adapter rotator, i.e. varying the parameter {\tt ADA.ABSROT}. If the gradient originates from outside the instrument, it should rotate with the instrument. For this test, we used the observations with id 7 in Tab.~\ref{tab:observations}. They have been obtained by setting the telescope into a non-guiding state (staring at zenith). This required some tricks on the telescope control software (TCS) side since the telescope normally ignores chopping commands when in non-guiding state. The Telescope and Instrument Operator (TIO) first put the telescope in non-guiding/non-tracking state and then modified an entry in the TCS online database (OLDB) to pretend that we were actually guiding. We also had to modify the database to pretend that the active optics loop was closed, which in fact was also not true. However, an active optics correction had been performed shortly before the observations and the active optics was running in open loop. Since we did not image any object, but just looked at the emission from sky and telescope, there were no requirements on image quality. These modifications were necessary because only then would the VISIR template run through (otherwise it would wait for the TCS to be brought in guiding state / the active optics loop to be closed).

\begin{figure}[ht]
\centering
\includegraphics[width=0.8\columnwidth]{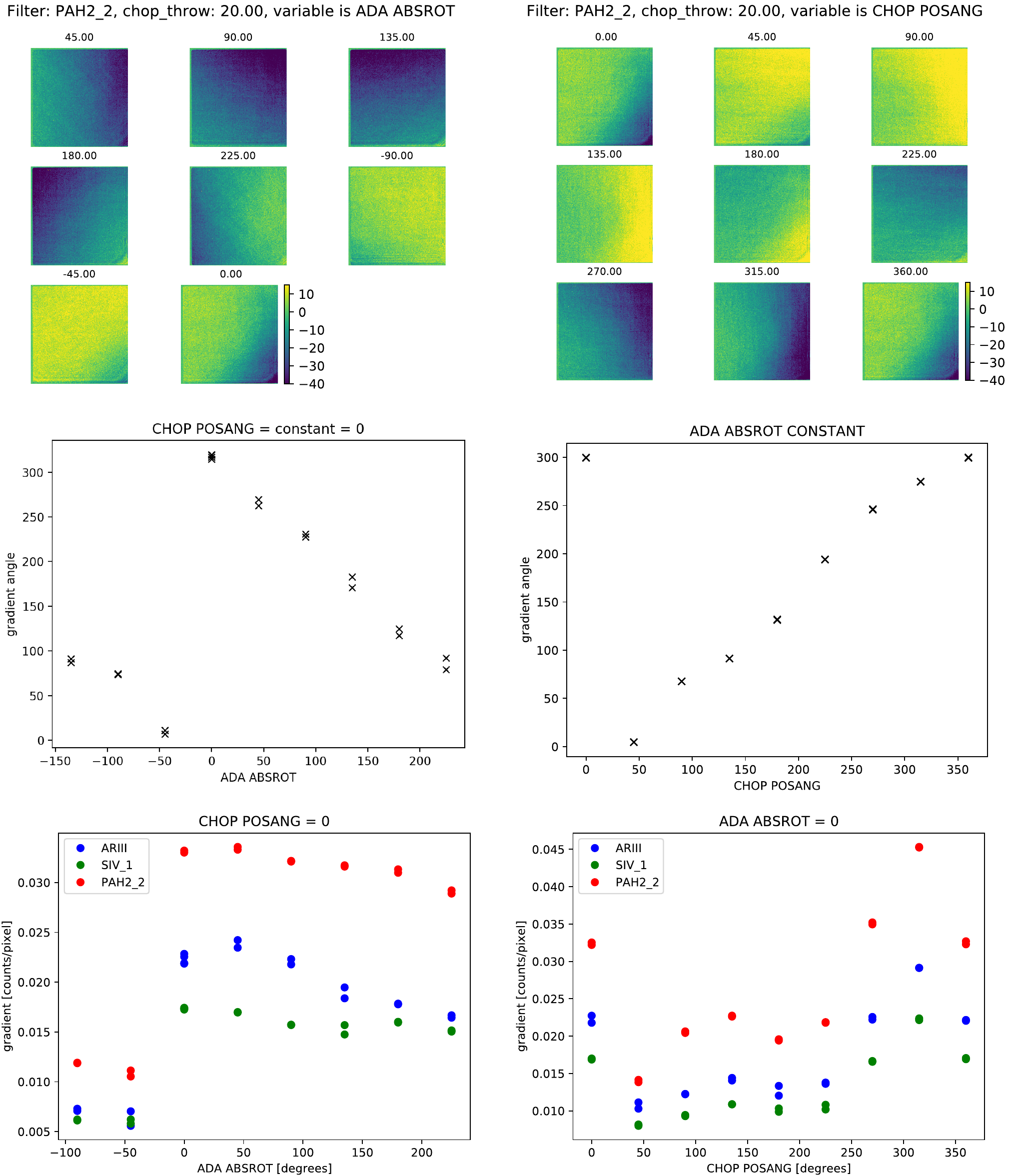}
\caption{\label{fig:grad_rotation}Sequence of chop-difference images (and analysis thereof) in the filter {\tt PAH2\_2}, obtained from rotating the instrument about the optical axis using the Cassegrain adapter-rotator ({\tt ADA ABSROT}, left column) and by ``rotating'' the chopping position angle ({\tt CHOP POSANG}, right column), respectively. Top row: Chop-difference images at indicated {\tt ADA ABSROT} and {\tt CHOP POSANG} values; middle row: fitted gradient angle vs. rotation angle; bottom row: absolute value of the gradient vs. rotation. At each derotator angle and at each chopping angle, two observations were taken leading to two data points per angle.}
\end{figure}

In this state, we rotated the instrument in 45 degree steps using the Cassegrain adapter rotator. We performed this test three times with the filters {\tt ARIII} ($\approx$ 9 $\mu$m), {\tt PAH2\_2}  ($\approx$ 12 $\mu$m) and {\tt SIV\_1} ($\approx$ 10.6 $\mu$m) and show the resulting sequence of chop difference images for the {\tt PAH2\_2} filter in Fig.~\ref{fig:grad_rotation} (left column). The gradient rotates counter-clockwise with the instrument as expected if the gradient is fixed in the telescope frame. The result in the other two filters is very similar, but a bit noisier due to their reduced bandwidth compared to the {\tt PAH2\_2} filter. The median value of the gradient is $0.0315 \pm 0.0089$ counts/pixel in the {\tt PAH2\_2} filter and scales approximately linearly with filter bandwidth, e.g. it is $0.01571 \pm 0.0043$ counts/pixel in the {\tt SIV\_1} filter which has approximately half the bandwidth of the {\tt PAH2\_2} filter. 

Except for the two negative rotation angles where the gradient is very low for reasons that we do not yet understand, the gradient does not change much in value. From this test we conclude that the gradient is fixed in the telescope frame, and not an instrumental property.

\subsubsection{Gradient behaviour under chopping angle rotation}

A similar effect is seen when changing the orientation of chopping but keeping the instrument rotation constant. This is shown for the {\tt PAH2\_2} filter in Fig.~\ref{fig:grad_rotation} (right column). The gradient rotates clockwise with the chopping position angle. The variation in amplitude is significantly larger than when rotating the instrument, i.e. the intrinsic gradient amplitude depends on the direction of chopping.

\subsection{The origin of the VISIR gradient}

\begin{figure}[ht]
\centering
\includegraphics[width=0.8\columnwidth]{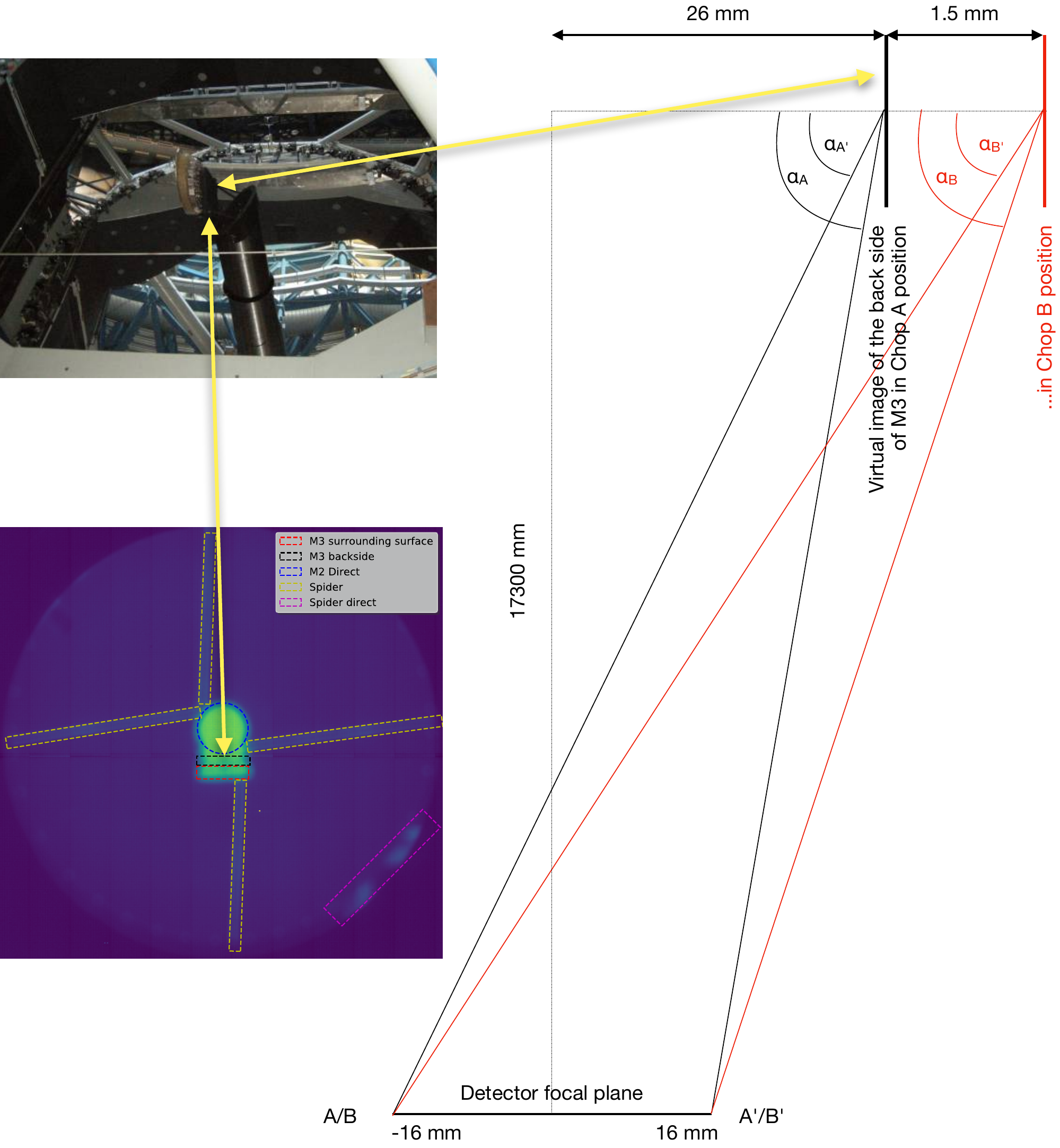}
\caption{\label{fig:m3_model}Top left: visual image of the M3 tower in VLT UT3 with M3 in its parked position (as for VISIR observations in the Cassegrain focus; picture courtesy H.-U. K\"aufl); bottom left: The M3 mirror and its support structure can be clearly seen in this VISIR pupil image taken with the {\tt ARIII} filter as well as a number of other structures; right: sketch to estimate the differential contribution from the M3 back side during chopping (see text for explanation).}
\end{figure}

When looking at a VISIR pupil image (test 1 from Tab.~\ref{tab:observations}; Fig.~\ref{fig:m3_model}) one immediately notices a number of emitters which we quantify here for the {\tt ARIII} filter: The spiders contribute about 3.7 \%, M2 contributes 11.1 \% and the M3 backside and surrounding surfaces contribute 2.4 \% and 2.9 \% respectively.

Since the gradient rotates together with the telescope, since it has a dependency on chop direction, and since the only asymmetric warm emitter in the pupil is the M3 structure, we suspect that this is the structure that is responsible for producing the gradient in chop difference images. More precisely, we expect that in particular the black back side of the M3 structure should lead to a measurable differential signal in the chop-difference frame. We note that thermal-IR instruments at telescopes that mask warm elements in the pupil in one way or another like GTC/CanariCam or Subaru/COMICS (see below) do not see such a gradient. Furthermore, the observation that the gradient rotates with instrument angle (and also with chopping position angle) with a 360 degree symmetry, excludes the spiders from which we would expect a 90 degree symmetry.

Let us therefore recall that the gradient is of order 20 counts (in the {\tt ARIII} filter, such as the pupil image used here) over the whole frame, and that the unchopped frame typically has of order 10,000 counts (bias-subtracted) of which we estimate that 2.4\% (240) originate from the M3 structure. If we attribute the entire gradient to the structure that produces these counts, then chopping could lead to a gradient of 20/240 $\sim$ 0.083 in the chop-difference frame across the entire detector.

We can model the projection effect that chopping produces in a simplified system that only consists of the VISIR detector and the M3 back side where VISIR sees it, i.e. a virtual image of M3, 17.3m above the Cassegrain focus. M3 is de-magnified due to the hyperbolic M2 mirror and appears 26mm off the optical axis and chopping offsets it by another 1.5mm (from our own optical calculations with numbers from the VLT white book\cite{vltwhite1998}). The detector itself is 32 x 32 mm in size. Under the assumption that the M3 back side emits at 100 \% emissivity and like a Lambertian emitter, we calculate the flux in arbitrary units that we get at the left edge of the detector in chop A and B positions as $\cos\alpha_A$ and $\cos\alpha_B$, respectively (see the sketch in Fig.~\ref{fig:m3_model}) and similarly for all other pixels on the detector. We can thus create simulated (1D) ``raw'' images in chop A and chop B state, img$_A$ and img$_B$, and also a simulated chop-difference images as img$_A$ - img$_B$. In order to compare this simulated chop-difference frame to observations, we normalise it by img$_A$ and arrive at a relative gradient of approximately 0.1 across the detector, very close to the contribution we expect from looking at the emitters in the pupil plane. We also investigated this effect further using a ray tracing model and confirm that this projection effect is very likely the origin of the gradient. This model will be presented in an upcoming article by Ioannis Politopoulos et al.

\subsection{Spider masking experiments with CanariCam}

To further test the connection between structures in the thermal-infrared pupil and chop-difference frames, we also obtained a number of engineering observations with CanariCam at the GTC. It is different from VISIR in a number of ways that make a comparison particularly relevant for the ELT: CanariCam has been installed in a Nasmyth focus and in a Folded-Cassegrain focus, always after the a reflection off the tertiary mirror M3, with similarities to the future METIS. In addition, GTC has a tessellated mirror (like the ELT will have).

\begin{figure}[ht]
\centering
\includegraphics[width=0.8\columnwidth]{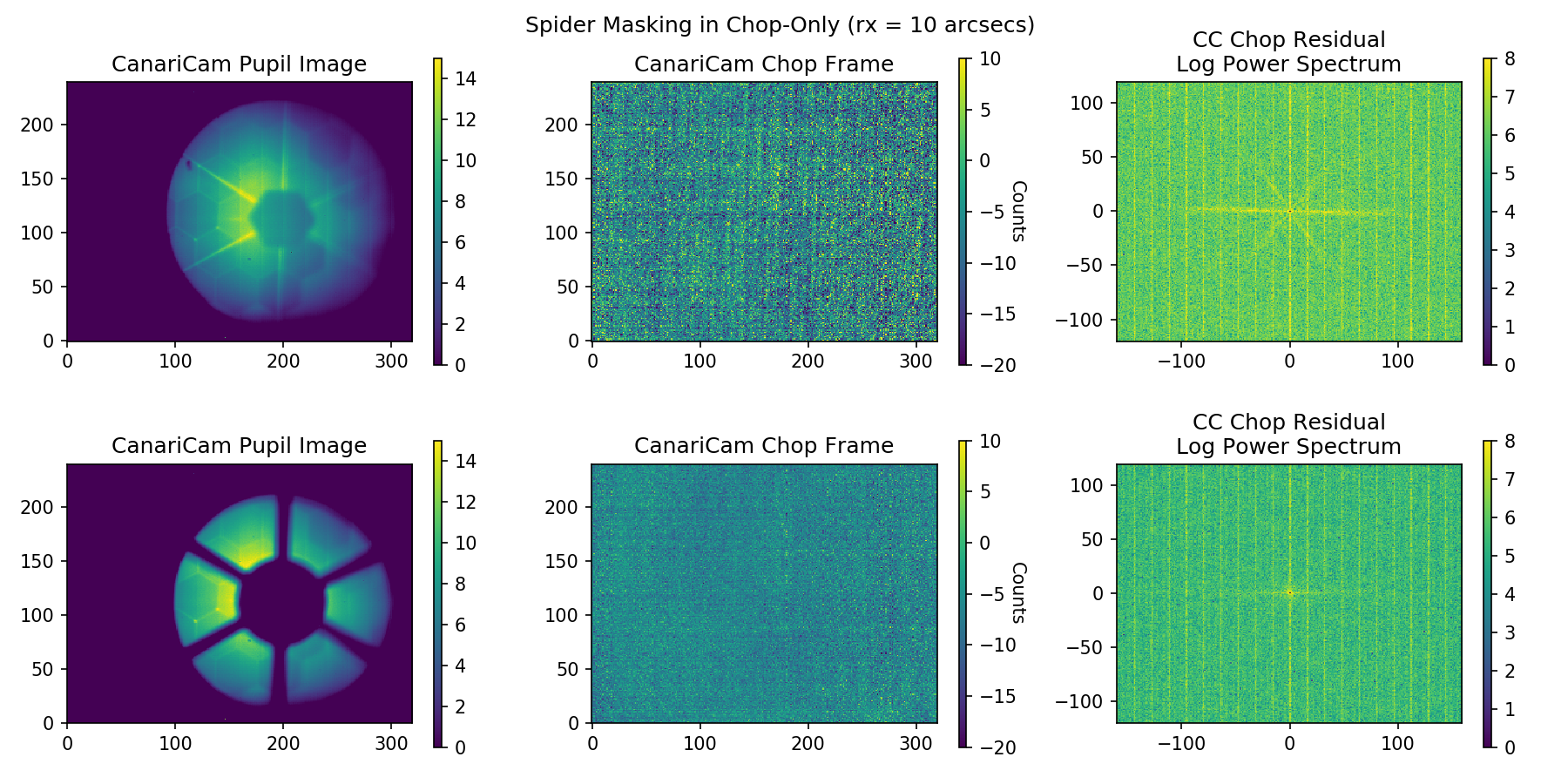}
\caption{\label{fig:CC_spider_rx}Spider-masking experiments with CanariCam. From left to right we show the (1) pupil image, (2) the chop-residual frame and the (3) power spectrum for the two experiments: top panels with the standard inscribed circular pupil stop (CIRC1), and bottom panels with the “Pie-Wedge” pupil stop that masks the spiders and the central area of the pupil. In both cases, the chop-throw was 10 arcsec along the $x$ axis at 1.6Hz with the ArIII filter under high background conditions (twilight and cirrus), and the telescope staring at the same sky position. The pupil mask removes most of the chop residual structure.}
\end{figure}

\begin{figure}[ht]
\centering
\includegraphics[width=0.8\columnwidth]{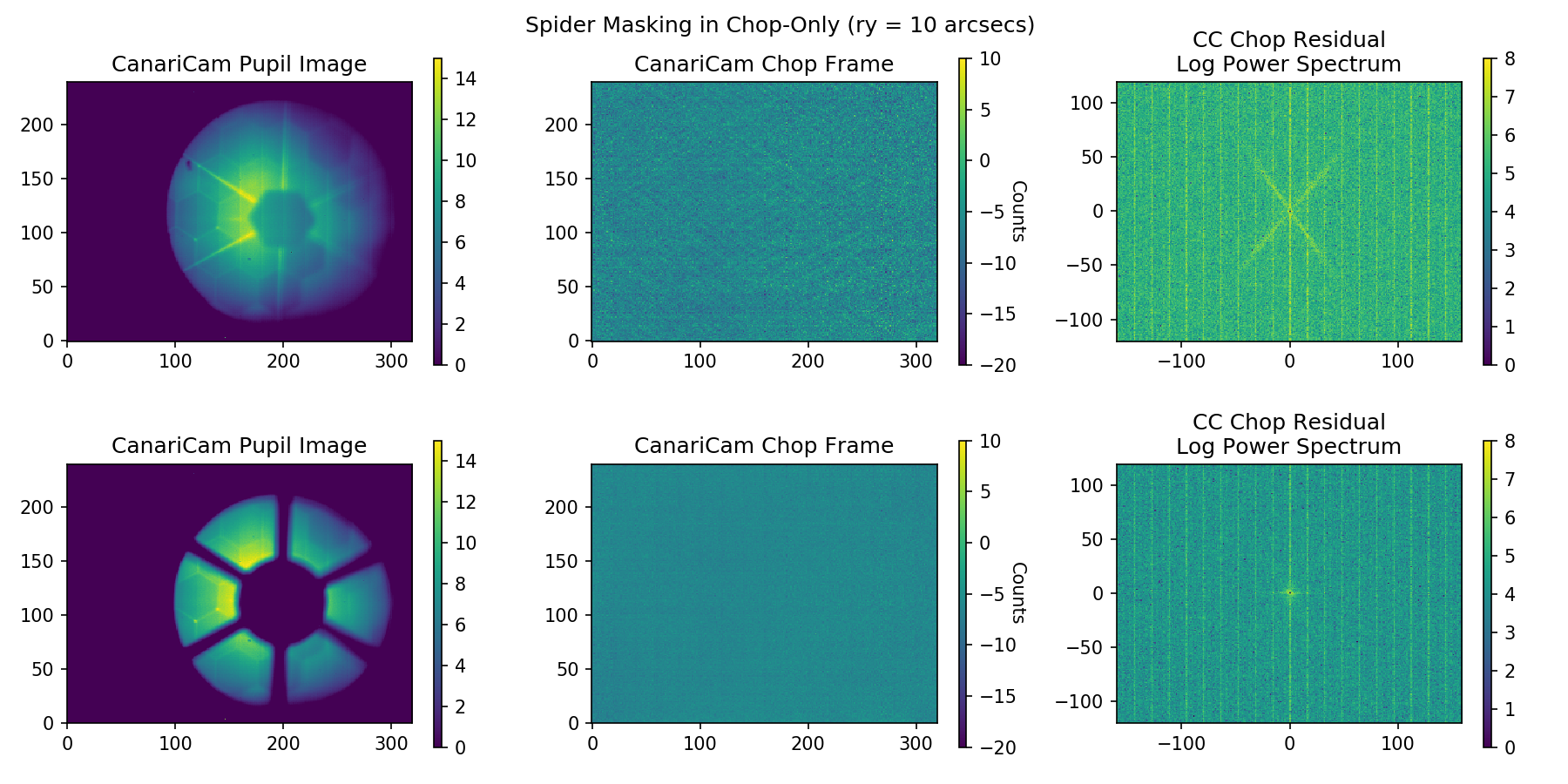}
\caption{\label{fig:CC_spider_ry}Spider-masking experiments with CanariCam. Same experiment as described in Fig.~\ref{fig:CC_spider_rx}, but the chopping direction is along the $y$ axis. Due to the different symmetry, one of the spiders is not seen in the power spectrum, demonstrating that the chop direction is relevant. In this case, the pupil mask also removes most of the structure.}
\end{figure}

We present in Figs.~\ref{fig:CC_spider_rx} and \ref{fig:CC_spider_ry} two sets of pupil images and chop-difference frames with CanariCam each, in which we study the impact of the chopping direction and of masking the spiders in the cold pupil (see the figures for more description). When the chopping direction is not aligned with the spider arms, a high-frequency noise can be seen in the chop-difference frame (Fig.~\ref{fig:CC_spider_rx}, top panels). The Fourier Transform image reveals that this noise is caused by the six spider arms in the warm pupil. While this noise will normally be removed through nod subtraction, it can lead to significant residuals in the chop-nod subtracted image if the nodding cycle is too slow compared to the field rotation. This has been identified earlier in VLT/VISIR as the origin of what was there called the ``fence effect''\cite{pantin2010}.

Blocking the spiders using a pupil mask (Fig.~\ref{fig:CC_spider_rx}, lower panel) removes this structure, leaving only a number of vertical stripes in the Fourier Transform image that originate from the distinct bias levels of the various detector readout columns. As Fig.~\ref{fig:CC_spider_ry} shows, a part of the ``spider noise'' can also be suppressed by choosing the chopping direction to coincide with one of the spider arms.

We conclude that the use of a pupil mask improves the chop residuals substantially by removing unwanted thermal radiation. For CanariCam, this does not mean that one always needs to observe in pupil-stabilized mode, by the way. CanariCam, unlike VISIR or METIS, can rotate its pupil mask to follow the pupil rotation and hence block static structures in the pupil in normal field-tracking observations. While the use of the pupil masks in CanariCam was mainly designed for the (pupil-tracking) coronographic mode (which was never commissioned), it could be used in a more general way, even for spectroscopy or faint extended sources using normal field tracking. In this case, it is uncertain if the chop direction should be fixed to the pupil (as was used in the experiments, which were done in engineering mode) or if one could just chop in sky coordinates since the pupil features are masked-out.

%%
%% +-+-+-+-+-+-+-+-+-+-+-+-+-+-+-+-+-+-+-+-+-+-+-+-+-+-+-+-+-+-+-+-+-+-+-+-+-+
%%

\section{DISCUSSION and CONCLUSIONS}

\subsection{Comparison of VISIR and CanariCam chopping residuals}

\begin{figure}[ht]
\centering
\includegraphics[width=\columnwidth]{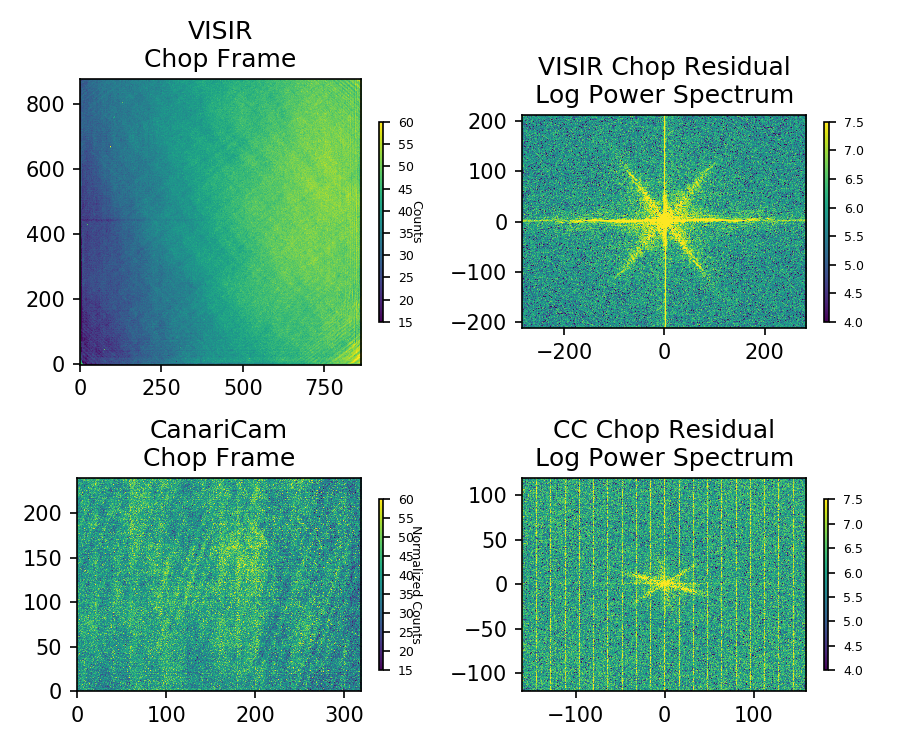}
\caption{\label{fig:chop_diff_compare}Comparison between the VLT/VISIR (top panels) and GTC/CanariCam (bottom panels) chop residual frames (left panels) and their power spectrum features in decimal log scale (right panels).
The VISIR setup was: {\tt PAH2\_2} filter; chop throw 10 arcsec at 3.8 Hz; 300 seconds total integration. Fixed rotator on blank sky.
The CanariCam setup was: {\tt Si2-8.7um} filter; chop throw 10 arcsec at 1.93 Hz; 240 seconds total integration. Fixed rotator on blank sky.
The CanariCam detector counts were normalized to the median level of VISIR, and the same scales were used for both instruments.  While the gradient is the dominant feature seen in the VLT/VISIR chop residual frame, GTC/CanariCam is dominated by the emission of the spiders and detector artifacts. The spiders and some detector artifacts can also be seen in the VISIR power spectrum.  Due to the different pixel scale (CanariCam 0.08 arcsec/pix vs VISIR 0.045 arcsec/pix), the shown frequency (1/pix) range has been adjusted.}
\end{figure}

We show a comparison of the VISIR and CanariCam chop-difference frames in Fig.~\ref{fig:chop_diff_compare}. As we have noted before, the VISIR chop-difference frame is dominated by the gradient and some artefacts in the lower right corner, which could be diffraction fringes from the field mask. However, the features produced by the 4-arms spider can be clearly seen in the Fourier Transform image.

CanariCam is dominated by a similar feature produced by its 6-arms spider along with detector readout artifacts, especially the 20 pixels wide channels. A higher chop frequency would likely reduce this detector noise, but CanariCam chose to sacrifice background sampling to increase duty-cycle since this approach yielded better sensitivity.

\subsection{Warm emitters in different telescope pupils}

\begin{figure}[ht]
\centering
\includegraphics[width=\columnwidth]{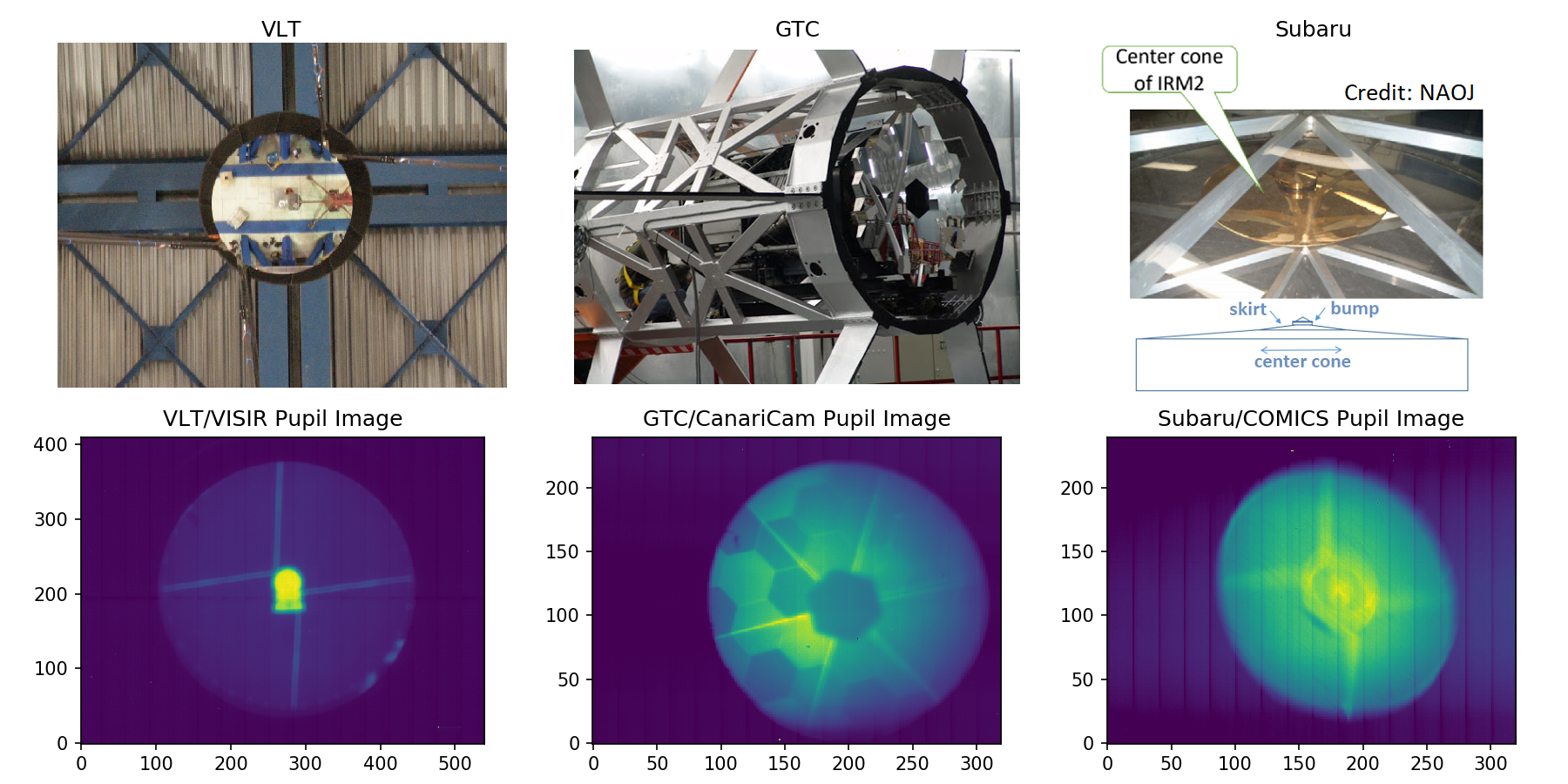}
\caption{\label{fig:warm_pupils}Different secondary mirror designs and their respective thermal-IR pupil images. Top panels, from left to right: (1) VLT/UT3 secondary mirror in its infrared configuration as seen from the Cassegrain port (picture courtesy of H.-U. K\"aufl); (2) GTC with its undersized secondary mirror matching the shape of M1, with an aperture in the center to suppress the thermal emission coming from the instrument port and the tertiary mirror; (3) Subaru's IR-optimized exchangeable secondary mirror (IRM2) with a center-cone designed to reject the thermal emission coming from the Cassegrain port. Bottom panels, from left to right: thermal-infrared pupil imaging frames taken with the instruments VISIR, CanariCam and COMICS, respectively.}
\end{figure}

For large telescopes at high-altitude sites with low typical columns of precipitable water vapour (PWV), the background level in the mid-IR is dominated by the telescope's own thermal emission.  Different design approaches have been followed over the years to minimize this contribution by (1) reducing the telescope emissivity (e.g., using silver or gold-coated mirrors) and/or (2) rejecting this radiation and preventing it from reaching the infrared instrumentation (see Fig.~\ref{fig:warm_pupils}).
It has been discussed that the source of the gradient seen on VISIR is the tertiary mirror in its park position.  To reject this emission and that coming from the instrument port as well, GTC and Subaru designed IR-optimized secondary mirrors. GTC adopted a solution with a central hole, with free view to the sky, matching the Cassegrain and the M3 projected size, whereas Subaru uses an exchangeable M2 and its IR-optimized unit has a center-cone to diverge the emission from the instrument port out of the pupil.  Subaru also adopted additional techniques like the use of IR mirrors to hide the support structure of the M2 and the M3 spiders, reaching a very low overall telescope emissivity of $\sim$ 5\% (priv. comm. Subaru team).

Since METIS will use a pupil mask that removes the emission from both the ELT spiders and the central obscuration, it should not show any of these known structures in its chop-difference frames. One relevant and challenging factor for ELTs will be the cleanliness of their segmented primary mirrors.  It must be noted that the GTC/CanariCam pupil image shown in Fig.~\ref{fig:warm_pupils} was taken in November 2020 when the more emissive segments had been in the queue for immediate recoating. Due the COVID19 pandemic, the segments cleaning and recoating schedule was heavily impacted by the lockdown and further restrictions, however.

\subsection{Conclusions}
Using dedicated test observations at both VLT/VISIR and GTC/CanariCam, we have identified three different possible origins of structures in chop-difference frames of ground-based thermal-infrared observations:

\begin{enumerate}
    \item After the exchange of the VISIR {\bf entrance window}, the chop-difference residuals changed systematically, from which we conclude that the old entrance window of VISIR must have produced the ``ripples'' seen in earlier observations. For ELT/METIS, the quality and cleanliness of the entrance window will be of even higher importance since both the internal chopper and de-rotator will move the footprint of the beam across the entrance window (even for pupil-tracked observations).
    \item The gradient in VISIR chop-difference images has been studied by varying the instrument rotation angle as well as the chopping position angle. Its absolute value has further been analysed using a simple geometrical model. We conclude that the {\bf backside of the folded M3 mirror} in its parking position is responsible for the gradient in chop-difference images in VISIR.
    \item High-frequency noise in chop-difference frames of both VLT/VISIR and GTC/CanariCam has been analysed using Fourier Transforms as well as pupil masking experiments. The noise originates from the {\bf spiders} and can be suppressed partly by choosing an optimal chopping direction, and completely by blocking the spider emission in the cold pupil of the instrument.
\end{enumerate}

While some of these effects may be specific for the instruments that we studied, we believe that we can draw the general conclusion from this work that warm emitters must be masked in the cold pupil of the instrument for optimal performance of ground-based thermal-infrared instruments. This requires either a rotating pupil stop (as in CanariCam) or pupil-stabilised observations (as planned for METIS) with field de-rotation in the post-processing pipeline. In such observations, we expect that the remaining chopping residuals will be dominated by (static or only slowly varying) imperfections in various elements along the optical path which can be removed with nodding offsets of much slower frequency (and therefore higher observing efficiency) than in current ground-based thermal infrared instruments.

%%
%% +-+-+-+-+-+-+-+-+-+-+-+-+-+-+-+-+-+-+-+-+-+-+-+-+-+-+-+-+-+-+-+-+-+-+-+-+-+
%%

\acknowledgments % equivalent to \section*{ACKNOWLEDGMENTS}       
The authors would like to acknowledge the Paranal and GTC/CanariCam teams for their excellent support in obtaining these (often non-standard) observations. The authors would also like to acknowledge the Lorentz center workshop ``The next generation of thermal-IR Astronomy'' (12-16 November 2018) and the ESO workshop ``Ground-based thermal infrared astronomy -- past, present and future'' (12-16 October 2020). Discussions and contacts at these workshops have been instrumental for the compilation of this article. In particular we would like to thank the Subaru team members Honda-san and Miyata-san for sharing their expertise with us.

% https://www.lorentzcenter.nl/the-next-generation-of-thermal-ir-astronomy-how-can-we-reach-the-photon-noise-limit.html
% https://www.eso.org/sci/meetings/2020/IR2020.html

This work is based on observations at the European Southern Observatory VLT, Calibration Programme ID 4101.L-0802 and on observations made with the Gran Telescopio Canarias (GTC), installed in the Spanish Observatorio del Roque de los Muchachos of the Instituto de Astrof\'isica de Canarias, in the island of La Palma.

%%
%% +-+-+-+-+-+-+-+-+-+-+-+-+-+-+-+-+-+-+-+-+-+-+-+-+-+-+-+-+-+-+-+-+-+-+-+-+-+
%%

% References
\bibliography{report} % bibliography data in report.bib
\bibliographystyle{spiebib} % makes bibtex use spiebib.bst

\end{document}